\documentclass[aps,prl,reprint,showpacs,amsmath,amssymb]{revtex4-1}

\usepackage{graphicx}
\usepackage{dcolumn}
\usepackage{bm}
\usepackage{color}
\usepackage{pifont}

\begin{document}

\title{The photonic trumpet: An efficient, broadband interface between a solid-state quantum emitter and a Gaussian beam}

\author{Mathieu Munsch}
\author{Nitin S. Malik}
\author{Jo{\"e}l Bleuse}
\author{Emmanuel Dupuy}
\author{Adrien Delga}
\author{Jean-Michel G\'erard}
\author{Julien Claudon}
\email{julien.claudon@cea.fr}
\affiliation{CEA-CNRS-UJF group "Nanophysique et Semiconducteurs", CEA, INAC, SP2M, F-38054 Grenoble, France}
\author{Niels Gregersen}
\author{Jesper M{\o}rk}
\affiliation{Department of Photonics Engineering, DTU Fotonik, Technical University of Denmark, Building 343, 2800 Kongens Lyngby, Denmark}

\begin{abstract} We introduce the photonic trumpet, a dielectric structure which ensures a nearly perfect coupling between an embedded quantum light source and a Gaussian free-space beam. A photonic trumpet exploits both the broadband spontaneous emission control provided by a single-mode photonic wire and the adiabatic expansion of this mode within a conical taper. Numerical simulations highlight the outstanding performance and robustness of this concept. As a first application in the field of quantum optics, we report the realisation of an ultra-bright single-photon source. The device, a GaAs photonic trumpet containing few InAs quantum dots, demonstrates a first-lens external efficiency of $0.75 \pm 0.1$.
\end{abstract}

\date{\today}
\maketitle


Feeding the spontaneous emission (SE) of a quantum emitter into a controlled optical channel represents a major challenge in the context of quantum communication, photonic quantum information processing and metrology~\cite{Shields_NatPhot_07}. In this perspective, optical waveguides ensuring an efficient SE control~\cite{Viasnoff-Schwoob_PRL_05,Lund-Hansen_PRL_08,Quan_PRA_09,Akimov_Nature_07,Claudon_NatPhot_10,Babinec_NatNano_10} have recently emerged as a promizing alternative to the traditionnal microcavity-based approach~\cite{Gerard_TAP_03,Vahala_Nature_03,Reithmaier_SST_08}. Their broad operation bandwidth and the absence of resonant photon recirculation are favourable to the scalable realization of bright quantum light sources. Moreover, waveguides embedding artificial atoms with various level schemes have been predicted to mediate strong non-linear interactions at the single-photon level~\cite{Chang_NatPhys_07,Roy_PRL_11,Zheng_PRL_11,Kolchin_PRL_11,Valente_PRA_12}, with applications to photonic quantum logic. Among the investigated platforms~\cite{Viasnoff-Schwoob_PRL_05,Lund-Hansen_PRL_08,Quan_PRA_09,Akimov_Nature_07}, fiber-like photonic wires present appealing performances~\cite{Claudon_NatPhot_10,Babinec_NatNano_10,Heinrich_APL_10,Reimer_NatComm_11}. These vertical cylinders, made of a high-index dielectric material, are particularly well suited to the far-field collection of light. When they embed an emitter with a transverse dipole (such as a self-assembled quantum dot), they offer a very efficient SE control~\cite{Bleuse_PRL_11,Munsch_PRL_12,Bulgarini_APL_12} combined with negligible optical losses. Furthermore, the integration of a metal-dielectric mirror below the wire~\cite{Friedler_OL_08} and a needle-like tapering of its upper end~\cite{Gregersen_OL_08} bring the light extraction efficiency close to unity~\cite{Friedler_OE_09,Claudon_NatPhot_10}.


Though directive, the far-field emission of these structures significantly differs from a Gaussian beam. A Gaussian emission is however highly desirable for a wide range of situations, including the efficient feeding of quantum light into a single mode fiber, crucial for long-range quantum communications. Free-space Gaussian beams are readily generated and manipulated by standard optical setups and can be used to address efficiently a localized emitter with a propagating photon~\cite{Rakher_PRL_09}. In the long run, quantum networks will require robust spin-photon interfaces with a controlled emission~\cite{Kimble_Nature_08}. Regarding this point, the performance of a needle taper is sensitive to minute geometrical details, thus severely compromising the device fabrication yield.
 
In this Letter, we introduce photonic trumpets - the result of the opposite tapering strategy - and show these structures offer a robust and clean Gaussian far-field emission. After presenting theoretical design guidelines, we demonstrate the first implementation of this broadband approach, through the demonstration of a very bright single-photon source, with a first-lens external efficiency of $0.75 \pm 0.1$. Photonic trumpets are also particularly well suited to the implementation of electrical contacts, opening a wealth of opportunities for the future developments of solid-state quantum optics.

\begin{figure}[t]
\includegraphics[width=0.4\textwidth]{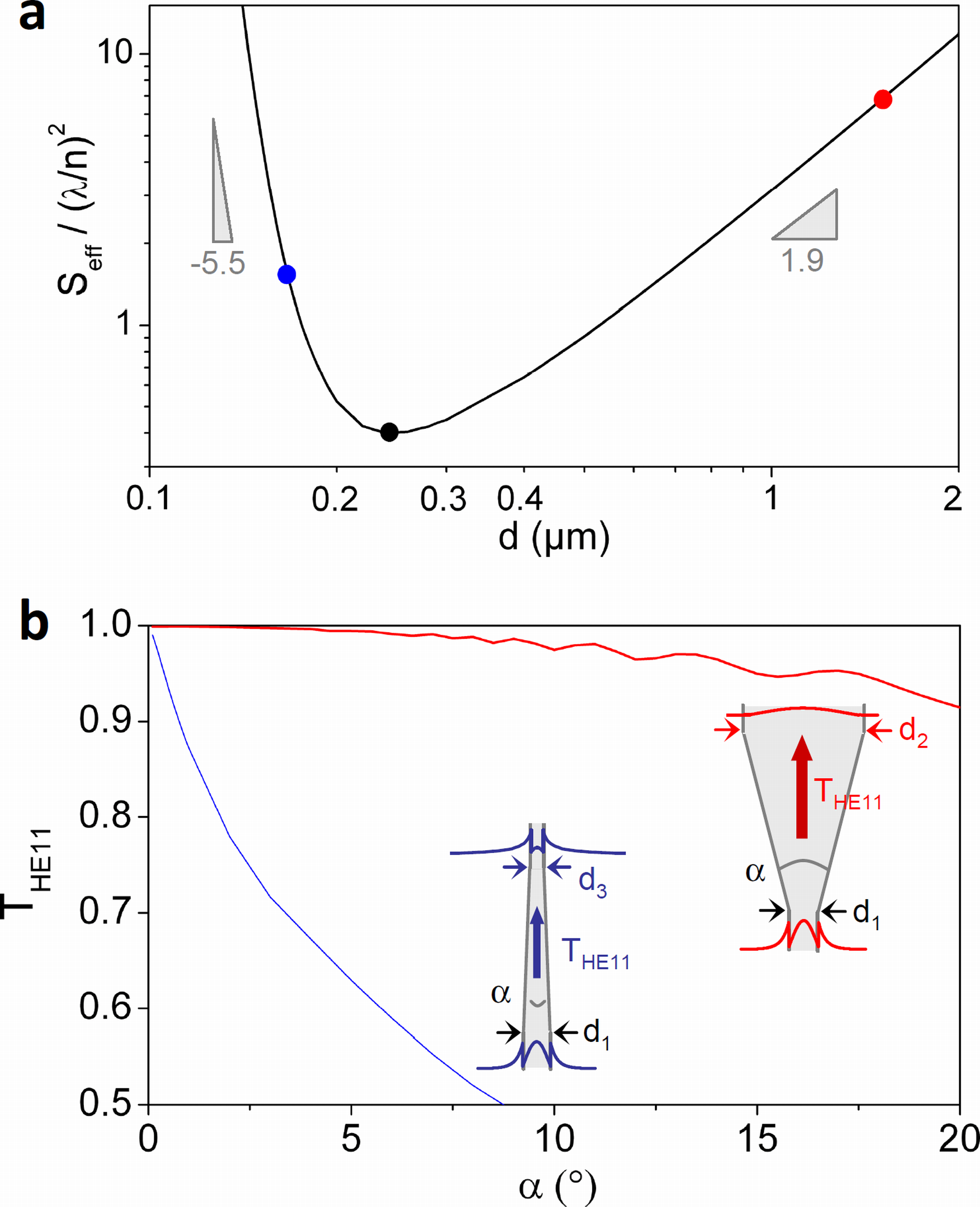}
\caption{(color online) {\bf Needle and trumpet tapers.} (b) Effective surface ($S_{\text{eff}}$) of the mode HE$_{11}$ guided by a cylindrical wire of diameter $d$ (double log scale, operation wavelength $\lambda = 950\: $nm). Starting from $d_1 = 240\: $nm, which optimizes the lateral confinement, $S_\text{eff}$ can be increased either by decreasing or increasing $d$, resulting in a needle or a  trumpet taper. (b) Modal transmission of HE$_{11}$ ($T_{\text{HE}_{11}}$), plotted against the tapering angle $\alpha$ for two representative tapers. For the photonic trumpet, $d_2$ is set to $1.5\: \mu$m; in the needle taper, $d_3=166\: $nm is chosen to ensure the same collection of the mode using a NA=0.8 lens. Typical electrical field profiles are also shown (amplitude of the discontinuous component). The dots appearing in (a) correspond to $d_1$, $d_2$ and $d_3$.}
\label{fig:figure1}
\end{figure}

The wire under consideration features a circular section of diameter $d$ and embeds a punctual, on-axis emitter with a transverse optical dipole (free-space wavelength $\lambda = 950\: $nm). It is made of GaAs, a high index material ($n=3.45$) and is immersed in a low index environment (air or vacuum). Such a dielectric waveguide always supports a fundamental guided mode (HE$_{11}$), whose lateral confinement is quantified by the effective surface:
$S_\text{eff} = \iint n(x,y)^2 \left| \mathbf{E}(x,y) \right|^2 \text{d}x \text{d}y / \big[ n(0,0)^2 \left| \mathbf{E}\right (0,0)|^2 \big],$
where $\mathbf{E}$ is the electric field amplitude. With such a normalization, the intensity of HE$_{11}$ zero-point fluctuations on the wire axis is proportional to $1/S_\text{eff}$. For $d_1 = 240\: $nm, the emitter is optimally coupled to the tightly confined guided mode (see Fig.~\ref{fig:figure1}(a)). Moreover, in this diameter range, the coupling to the 3D continuum of non-guided modes is strongly suppressed, thanks to a pronounced dielectric screening effect. As a consequence, the fraction $\beta$ of SE coupled to the HE$_{11}$ modes propagating upward and downward reaches $0.96$~\cite{Bleuse_PRL_11}. Furthermore, $\beta>0.9$ is maintained over a large wavelength range ($> 100\: $nm), a key asset of this 1D photonic system.

\begin{figure}[t]
\includegraphics[width=0.45\textwidth]{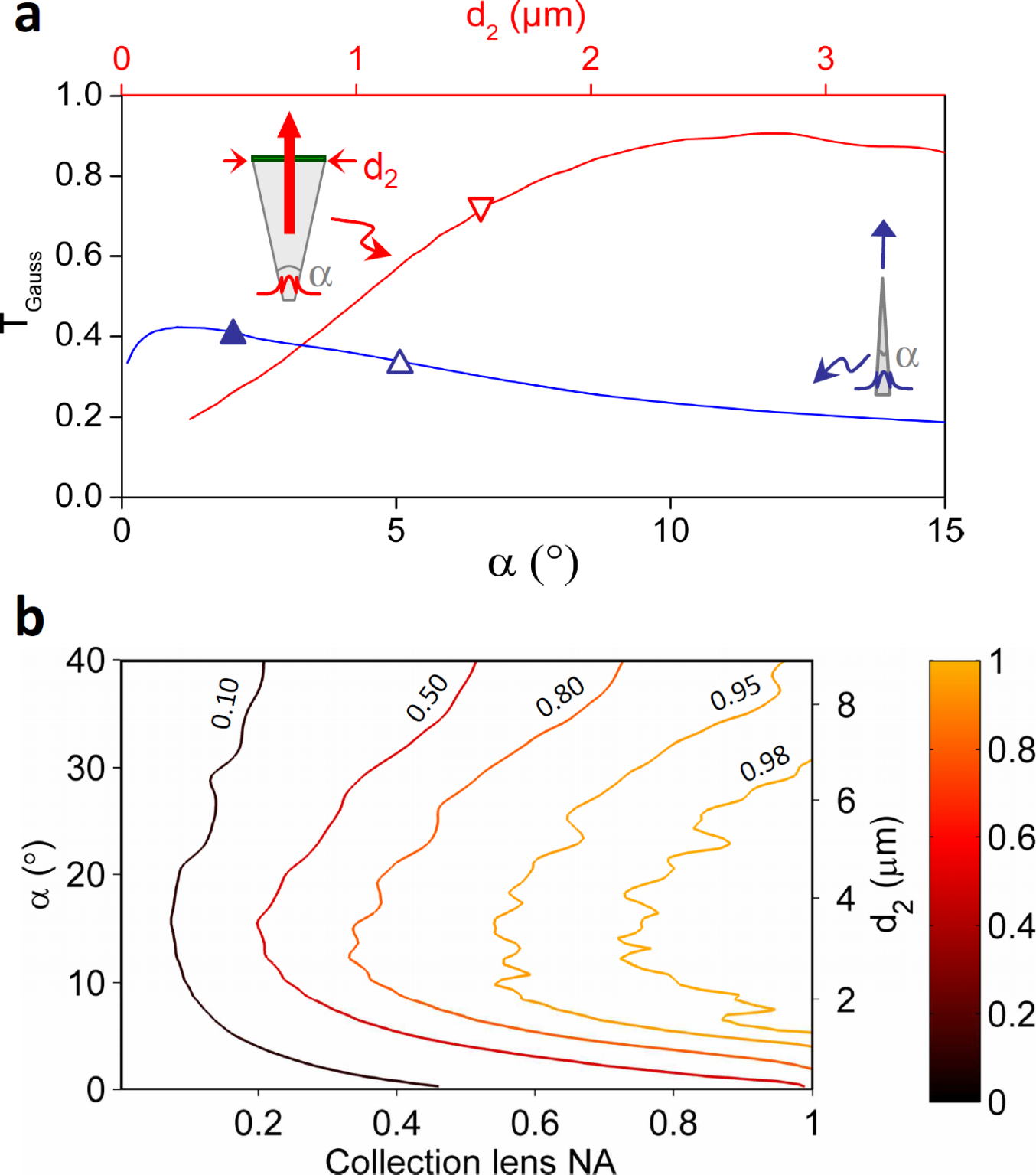}
\caption{(color online) {\bf Far-field emission of photonic trumpets.} (a) Calculated transmission $T_\text{Gauss}$ to a Gaussian beam using a NA=0.8 coupling lens. The evaluation is conducted for a $12\: \mu$m high photonic trumpet with various taper angle $\alpha$ (and thus different top diameter $d_2$). $T_\text{Gauss}$ for needle tapers is also shown. Experimental realisations: needle tapers {\tiny $\triangle$} (Ref.~\cite{Claudon_NatPhot_10}), $\blacktriangle$ (Ref.~\cite{Reimer_NatComm_11}) and photonic trumpet $\triangledown$ (this work). (b) Color map of $T_\text{tot}$, the total transmission of a photonic trumpet into a collection lens. $T_\text{tot}$ is plotted against the lens NA and $\alpha$.}
\label{fig:figure2}
\end{figure}

For a vast majority of applications, it is however desirable to optimize the coupling to standard free-space optics. Given the tight confinement of HE$_{11}$, a lateral expansion of the mode is then mandatory~\cite{Gregersen_OL_08}. As shown in Fig.~\ref{fig:figure1}(a), this can be achieved either through a decrease or an increase in $d$, resulting in needle-like and trumpet-like tapers. In both cases, we consider linear tapers (tapering angle $\alpha$) and first investigate $T_{\text{HE}_{11}}$, the modal transmission of HE$_{11}$ along a tapered section. Figure~\ref{fig:figure1}(b) shows $T_{\text{HE}_{11}}$ versus $\alpha$ for representative needle and trumpet tapers, calculated using the eigenmode expansion technique~\cite{Bienstman_OQE_01} with improved perfectly matched layers~\cite{Gregersen_OQE_08}. In a photonic trumpet, a nearly perfect adiabatic expansion of HE$_{11}$ is achieved for $\alpha < 5^{\circ}$, leading to $T_{\text{HE}_{11}}>0.994$. For $\alpha = 5^{\circ}$, the needle taper already suffers from significant non-adiabatic losses, which result in free-space emission before reaching the taper end. Qualitatively, the striking contrast between the two tapers can be understood by inspecting Fig.~\ref{fig:figure1}(a). For a needle taper, $S_\text{eff} \sim 1/d^{5.5}$ whereas the trumpet exhibits a $S_\text{eff}\sim d^{1.9}$ scaling law. Along the taper, the rate of change in diameter is governed by $\alpha$. A weaker dependence of $S_\text{eff}$ on $d$ thus implies slower changes in the mode profile during its propagation, which eases the adiabatic transformation of HE$_{11}$. In a trumpet with $\alpha > 5^{\circ}$, HE$_{11}$ experiences an increasing coupling to higher order guided modes. The propagation dynamics is then more complex, but $T_{\text{HE}_{11}}$ still exceeds $0.95$ for $\alpha$ as large as $15^{\circ}$. Such a tolerance is crucial from a practical point of view: it considerably alleviates the fabrication constraints and yields reproducible taper performances.


\begin{figure}[t]
\includegraphics[width=0.4\textwidth]{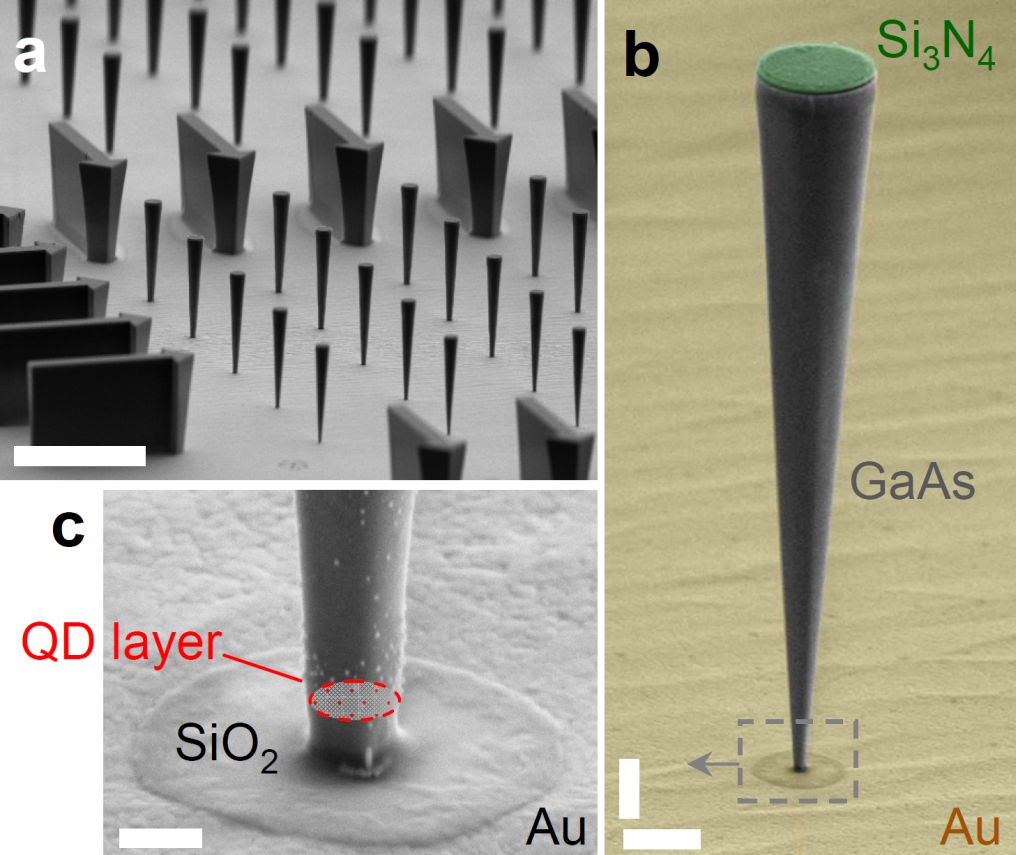}
\caption{(color online) {\bf Single-photon trumpets.} (a) Scanning electron microscope view of a large field of devices, illustrating the reproducibility of the fabrication process. Scale bar: $15\: \mu$m. (b) Zoom on a representative device (false colors); note the excellent control over the structure geometry. Vertical and horizontal scale bars: $1\: \mu$m. (c) Zoom on the connection between the trumpet and the integrated mirror. Scale bar: $200\: $nm.}
\label{fig:figure3}
\end{figure}

We now investigate the far-field emission of a $h=12\: \mu$m high photonic trumpet, whose top facet features a $\lambda/4$ anti-reflection coating. Since a clean Gaussian far-field emission is crucial for a wide range of applications, we first evaluate $T_\text{Gauss}$, the taper transmission to a Gaussian free-space mode using a NA=0.8 coupling lens. The calculation is conducted for various tapering angles, leading to different top facet diameters $d_2 = d_1 + 2h \tan (\alpha/2)$; the results are shown in Fig.~\ref{fig:figure2}(a). In the 'small' diameter range, $T_\text{Gauss}$ increases with $d_2$ to reach $0.91$ for $d_2 = 2.8\: \mu$m. Above this diameter, which corresponds to $\alpha > 12^{\circ}$, $T_\text{Gauss}$ undergoes a slight oscillating decrease due to the onset of mode conversion inside the taper. Slightly improved performance can be obtained with a higher taper: for $h=15\: \mu$m, a maximum $T_\text{Gauss}=0.93$ can be achieved. Again, photonic trumpets largely outperform even the sharpest needle tapers (Fig.~\ref{fig:figure2}(a)). This difference essentialy stems from the very favourable profile of HE$_{11}$ when it leaves the top facet of a trumpet.

The fraction $\epsilon_\text{Gauss}$ of SE coupled to the Gaussian mode can be further optimized through the integration of a gold-silica planar mirror below the trumpet. This mirror reflects the HE$_{11}$ mode propagating downward back into the wire~\cite{Friedler_OL_08}, with an amplitude modal reflectivity $r$. The reflection generates a standing wave pattern between the emitter and the mirror; locating the emitter on an electrical field antinode provides a boost to the SE rate by a factor of $(1+|r|)$ and maximizes the collection efficiency. In these conditions, $\epsilon_\text{Gauss}$ can be expressed as~\cite{Friedler_OE_09}:
\begin{equation}
\epsilon_\text{Gauss} = \frac{1}{2}  \frac{\beta(1+|r|)^2}{1+\beta |r|} T_\text{Gauss}.
\end{equation}
Using $\beta=0.96$, $|r|=0.95$~\cite{Friedler_OL_08} and $T_\text{Gauss}=0.93$, one obtains $\epsilon_\text{Gauss}=0.89$. Such a large value highlights the potential of photonic trumpets as a broadband and efficient interface between a quantum emitter and a Gaussian beam. 

Interestingly enough, for situations that do not impose any strong constraint on the far-field structure ({\it e.g.} free-space quantum key distribution), photonic trumpets can be operated deep into the non-adiabatic regime. Figure~\ref{fig:figure2}(b) shows the total taper transmission $T_\text{tot}$ into collection lenses with various NA. For NA = 0.8, $T_\text{tot} > 0.95$ is maintained for $\alpha$ as large as $32^{\circ}$. In that case, the constraints on $\alpha$ are thus further relaxed, which offers in particular the possibility to implement compact tapers.

Having discussed the potential of photonic trumpets, we now demonstrate the first implementation of this tapering strategy through the realisation of an on-demand, ultrabright single-photon source. The trumpets, made of GaAs, embed a single layer of self-assembled InAs quantum dots (QD) which are efficient and stable single-photon emitters. They are supported by a gold-silica planar mirror. A representative structure is shown in Fig.~\ref{fig:figure3}(b); it has been processed out of a planar sample grown by molecular beam epitaxy, using a top-down approach. The fabrication process of these high aspect-ratio structures begins with the deposition of the metal-dielectric mirror, followed by a flip-chip gluing on an host substrate. After removal of the growth wafer, the trumpets are defined with e-beam lithography and a carefully optimized dry etching step. The large SEM view in Fig.~\ref{fig:figure3}(a) illustrates the reproducibility of the fabrication; the zooms in Fig.~\ref{fig:figure3}(b) and (c) show the excellent control over the trumpet geometry, notably the connection to the planar mirror. The bottom part of the trumpet features a diameter in the $200-240\: $nm range. The structures are $12\: \mu$m high with $\alpha = 6.5^{\circ}$, which leads to a $1.55\: \mu$m top diameter. The QDs are located $110\: $nm above the mirror.



\begin{figure}[t]
\includegraphics[width=0.45\textwidth]{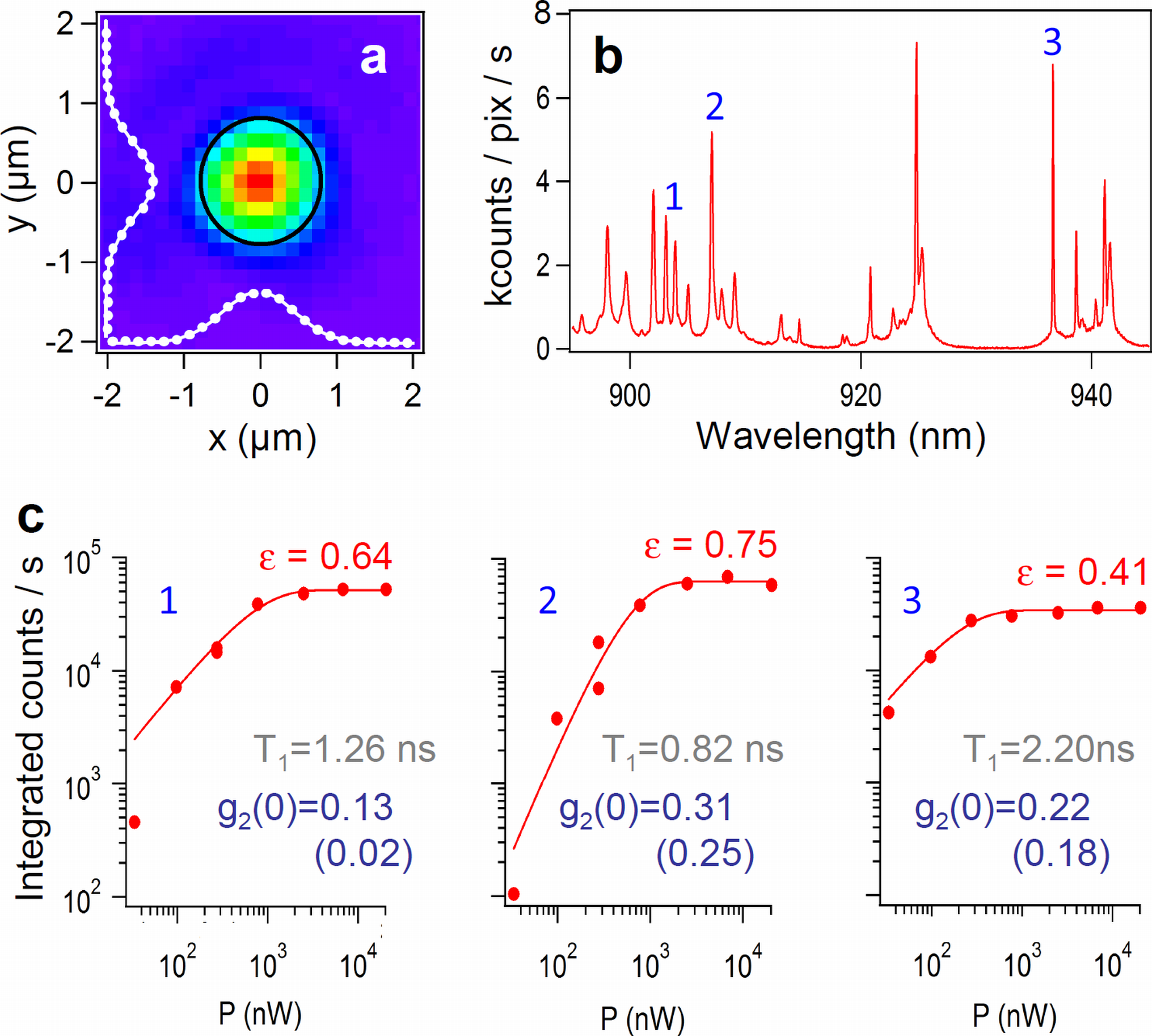}
\caption{(color online) {\bf Optical characterization.} (a) Spatial distribution of the intensity in the top facet plane. At the location of the QD, only HE$_{11}$ is guided by the structure and all the QDs contribute to the signal. Two diameter cut and their fit to a Gaussian profile are also shown. The circle represents the facet circumference. (b) Micro-photoluminescence spectrum, measured at $T=5\: $K for an excitation power $P=20\: \mu$W. (c) Spectrally integrated intensity of the lines 1, 2 and 3, extracted from a fit to a Lorentzian lineshape, as a function of $P$. The solid lines are theory, assuming an excitonic emission (lines 1 and 3) and biexctonic emission (line 2). For each transitions, the luminescence decay time $T_1$ and the intensity autocorrelation value $g_2(0)$ are also shown. The value in parenthesis excludes the background contribution (see Methods).}
\label{fig:figure4}
\end{figure}

The devices are operated at liquid helium temperature, using a micro-photoluminescence ($\mu$PL) setup equipped with a commercial microscope objective (NA=0.75). QD excitation is provided by a pulsed laser, tuned to the absorption continuum of the QD wetting layer, below the GaAs bandgap. Figure~\ref{fig:figure4}(a) shows the spatial distribution of the QD emission in the top facet plane: it presents a Gaussian shape, which describes satisfyingly HE$_{11}$ for this range of lateral confinement. The $\mu$PL spectrum, dispersed by a grating spectrometer and recorded with a CCD camera, features separated sharp lines, associated with QD excitonic transitions (Fig.~\ref{fig:figure4}(b)). In the following, we focus on three lines labelled 1, 2 and 3 ($\lambda=902.5\: $nm, $907.1\: $nm, $935.1\: $nm) associated with three different QDs. Their spectrally integrated intensities $I_\text{int}$, obtained from a fit to a Lorentzian lineshape, are plotted against the pump power in Fig.~\ref{fig:figure4}(c). Considering the low power dependence of $I_\text{int}$ and the measured transition decay time $T_1$, lines 1 and 3 are attributed to the recombination of an exciton and line 2 to the recombination of a bi-exciton. 

In each case, single-photon emission is assessed with a measurement of the intensity autocorrelation function $g_2(\tau) = \left< I(t) I(t+\tau) \right> / \left< I(t) \right>^2$, where the brackets represent a time averaging. The measurement is performed under pulsed excitation, using a Hanbury Brown and Twiss setup which employs two silicon avalanche photo diodes (time jitter: $300\: $ps). The raw values of $g_2(0)$ for lines 1-3, extracted from measurements with a spectral integration window of $140\: $pm, are indicated in Fig.~\ref{fig:figure4}(c). They are smaller than $0.5$, proving that the emission is dominated by the radiative recombination of a single electron-hole pair. Compared to our previous work~\cite{Claudon_NatPhot_10}, the higher dot density generates a sizeable luminescence background. Its contribution is removed in $g_2^{\star}(0)$, assuming statistical independence between the signal and a Poissonian background. The corresponding values appear between parenthesis in Fig.~\ref{fig:figure4}(c); in particular, line 1 exhibits $g_2^{\star}(0) = 0.02$, the signature of a pure single-photon emission.

The brightness of the source is quantified by the external efficiency $\epsilon$, defined as the probability to collect a single-photon in the first lens of the setup after an optical excitation pulse. The determination of $\epsilon$ requires a careful calibration of the setup, that we have conducted using a laser tuned to the QD emission wavelength as a reference, similarly to Ref.~\cite{Claudon_NatPhot_10}. The values that are given in the following are corrected from residual multiphoton events~\cite{Pelton_PRL_02}. When driven to saturation, line 2 exhibits $\epsilon = 0.75 \pm 0.1$. Such a large value exceeds the state-of-the-art for solid-state single-photon sources (QD-VCSEL microcavity~\cite{Strauf_NatPhot_07}, QD-needle like photonic wire~\cite{Claudon_NatPhot_10}, molecule-dielectric antenna~\cite{Lee_NatPhot_11}). Nevertheless, it remains smaller than the maximal theoretical value ($0.89$) calculated using the device dimensions. This small discrepancy is attributed to a non-optimal lateral positionning of the QD. Furthermore, lines 1 and 3 are also very bright ($\epsilon = 0.61$ and $0.41$), illustrating directly the high operation bandwidth of the device. Finally, efficiencies in the $0.5-0.6$ range where routinely obtained in other devices.

In the first realization reported here, the top facet diameter ($d_2= 1.55\: \mu$m) is already sufficient to ensure a large transmission to a Gaussian beam ($T_{\text{Gauss}} = 0.71$ for a NA=0.8 microscope objective). Compared to the best needle tapers, this already constitutes an improvement by a factor exceeding $1.6$. Future fabrication efforts will concentrate on the obtention of a wider top facet (through an increase of $h$ and/or $\alpha$) and should bring $T_{\text{Gauss}}$ close to ideality (see Fig.~\ref{fig:figure2}(b)). Moreover, a 'wide' top facet also leads to a very directive far-field emission, enabling the use of collection optics with a moderate NA. As an example, one can reach $T_{\text{Gauss}}=0.85$ with NA=0.4 for a trumpet characterized by $h = 15\: \mu$m and $d_2= 2.8\: \mu$m. Regarding advanced quantum light sources, the circular top facet is very convenient to implement a top electrode~\cite{Bockler_APL_08,Gregersen_OE_10}, which is desirable to provide an electrical charge injection in the QD~\cite{Yuan_Science_02}, or to tune its fine spectral properties with an electric field~\cite{Finley_PRB(R)_04,Bennett_NatPhys_10}.




In conclusion, photonic trumpets dramatically alleviate the fabrication constraints of tapered photonic nanowires, while offering a largely improved Gaussian far-field emission. The realization of an on-demand, ultrabright single-photon source represents a first implementation of this broadband approach. Considering the compatibility with the implementation of electrical contacts, we anticipate photonic trumpets will constitute a robust and generic platform for a wide range of solid-state quantum optics applications.

The authors acknowledge the support of the French 'Agence Nationale de la Recherche', under grant WIFO and the `Nanosciences aux limites de la Nano\'electronique'' Fundation, under grant STRONGCHIP. The sample fabrication has been performed in the `Plateforme technologique amont' and CEA L{\'e}ti/DOPT/SIONA clean rooms.


\begin{thebibliography}{35}%
\makeatletter
\providecommand \@ifxundefined [1]{%
 \@ifx{#1\undefined}
}%
\providecommand \@ifnum [1]{%
 \ifnum #1\expandafter \@firstoftwo
 \else \expandafter \@secondoftwo
 \fi
}%
\providecommand \@ifx [1]{%
 \ifx #1\expandafter \@firstoftwo
 \else \expandafter \@secondoftwo
 \fi
}%
\providecommand \natexlab [1]{#1}%
\providecommand \enquote  [1]{``#1''}%
\providecommand \bibnamefont  [1]{#1}%
\providecommand \bibfnamefont [1]{#1}%
\providecommand \citenamefont [1]{#1}%
\providecommand \href@noop [0]{\@secondoftwo}%
\providecommand \href [0]{\begingroup \@sanitize@url \@href}%
\providecommand \@href[1]{\@@startlink{#1}\@@href}%
\providecommand \@@href[1]{\endgroup#1\@@endlink}%
\providecommand \@sanitize@url [0]{\catcode `\\12\catcode `\$12\catcode
  `\&12\catcode `\#12\catcode `\^12\catcode `\_12\catcode `\%12\relax}%
\providecommand \@@startlink[1]{}%
\providecommand \@@endlink[0]{}%
\providecommand \url  [0]{\begingroup\@sanitize@url \@url }%
\providecommand \@url [1]{\endgroup\@href {#1}{\urlprefix }}%
\providecommand \urlprefix  [0]{URL }%
\providecommand \Eprint [0]{\href }%
\providecommand \doibase [0]{http://dx.doi.org/}%
\providecommand \selectlanguage [0]{\@gobble}%
\providecommand \bibinfo  [0]{\@secondoftwo}%
\providecommand \bibfield  [0]{\@secondoftwo}%
\providecommand \translation [1]{[#1]}%
\providecommand \BibitemOpen [0]{}%
\providecommand \bibitemStop [0]{}%
\providecommand \bibitemNoStop [0]{.\EOS\space}%
\providecommand \EOS [0]{\spacefactor3000\relax}%
\providecommand \BibitemShut  [1]{\csname bibitem#1\endcsname}%
\let\auto@bib@innerbib\@empty
\bibitem [{\citenamefont {Shields}(2007)}]{Shields_NatPhot_07}%
  \BibitemOpen
  \bibfield  {author} {\bibinfo {author} {\bibfnamefont {A.}~\bibnamefont
  {Shields}},\ }\href@noop {} {\bibfield  {journal} {\bibinfo  {journal}
  {Nature Photon.}\ }\textbf {\bibinfo {volume} {1}},\ \bibinfo {pages} {215}
  (\bibinfo {year} {2007})}\BibitemShut {NoStop}%
\bibitem [{\citenamefont {Viasnoff-Schwoob}\ \emph {et~al.}(2005)\citenamefont
  {Viasnoff-Schwoob}, \citenamefont {Weisbuch}, \citenamefont {Benisty},
  \citenamefont {Olivier}, \citenamefont {Varoutsis}, \citenamefont
  {Robert-Philip}, \citenamefont {Houdr{\'e}},\ and\ \citenamefont
  {Smith}}]{Viasnoff-Schwoob_PRL_05}%
  \BibitemOpen
  \bibfield  {author} {\bibinfo {author} {\bibfnamefont {E.}~\bibnamefont
  {Viasnoff-Schwoob}}, \bibinfo {author} {\bibfnamefont {C.}~\bibnamefont
  {Weisbuch}}, \bibinfo {author} {\bibfnamefont {H.}~\bibnamefont {Benisty}},
  \bibinfo {author} {\bibfnamefont {S.}~\bibnamefont {Olivier}}, \bibinfo
  {author} {\bibfnamefont {S.}~\bibnamefont {Varoutsis}}, \bibinfo {author}
  {\bibfnamefont {I.}~\bibnamefont {Robert-Philip}}, \bibinfo {author}
  {\bibfnamefont {R.}~\bibnamefont {Houdr{\'e}}}, \ and\ \bibinfo {author}
  {\bibfnamefont {C.~J.~M.}\ \bibnamefont {Smith}},\ }\href@noop {} {\bibfield
  {journal} {\bibinfo  {journal} {Phys. Rev. Lett.}\ }\textbf {\bibinfo
  {volume} {95}},\ \bibinfo {pages} {183901} (\bibinfo {year}
  {2005})}\BibitemShut {NoStop}%
\bibitem [{\citenamefont {Lund-Hansen}\ \emph {et~al.}(2008)\citenamefont
  {Lund-Hansen}, \citenamefont {Stobbe}, \citenamefont {Julsgaard},
  \citenamefont {Thyrrestrup}, \citenamefont {S{\"u}nner}, \citenamefont
  {Kamp}, \citenamefont {Forchel},\ and\ \citenamefont
  {Lodahl}}]{Lund-Hansen_PRL_08}%
  \BibitemOpen
  \bibfield  {author} {\bibinfo {author} {\bibfnamefont {T.}~\bibnamefont
  {Lund-Hansen}}, \bibinfo {author} {\bibfnamefont {S.}~\bibnamefont {Stobbe}},
  \bibinfo {author} {\bibfnamefont {B.}~\bibnamefont {Julsgaard}}, \bibinfo
  {author} {\bibfnamefont {H.}~\bibnamefont {Thyrrestrup}}, \bibinfo {author}
  {\bibfnamefont {T.}~\bibnamefont {S{\"u}nner}}, \bibinfo {author}
  {\bibfnamefont {M.}~\bibnamefont {Kamp}}, \bibinfo {author} {\bibfnamefont
  {A.}~\bibnamefont {Forchel}}, \ and\ \bibinfo {author} {\bibfnamefont
  {P.}~\bibnamefont {Lodahl}},\ }\href@noop {} {\bibfield  {journal} {\bibinfo
  {journal} {Phys. Rev. Lett.}\ }\textbf {\bibinfo {volume} {101}},\ \bibinfo
  {pages} {113903} (\bibinfo {year} {2008})}\BibitemShut {NoStop}%
\bibitem [{\citenamefont {Quan}\ \emph {et~al.}(2009)\citenamefont {Quan},
  \citenamefont {Bulu},\ and\ \citenamefont {Lon\v{c}ar}}]{Quan_PRA_09}%
  \BibitemOpen
  \bibfield  {author} {\bibinfo {author} {\bibfnamefont {Q.}~\bibnamefont
  {Quan}}, \bibinfo {author} {\bibfnamefont {I.}~\bibnamefont {Bulu}}, \ and\
  \bibinfo {author} {\bibfnamefont {M.}~\bibnamefont {Lon\v{c}ar}},\
  }\href@noop {} {\bibfield  {journal} {\bibinfo  {journal} {Phys. Rev. A}\
  }\textbf {\bibinfo {volume} {80}},\ \bibinfo {pages} {011810} (\bibinfo
  {year} {2009})}\BibitemShut {NoStop}%
\bibitem [{\citenamefont {Akimov}\ \emph {et~al.}(2007)\citenamefont {Akimov},
  \citenamefont {Mukherjee}, \citenamefont {Yu}, \citenamefont {Chang},
  \citenamefont {Zibrov}, \citenamefont {Hemmer}, \citenamefont {Park},\ and\
  \citenamefont {Lukin}}]{Akimov_Nature_07}%
  \BibitemOpen
  \bibfield  {author} {\bibinfo {author} {\bibfnamefont {A.}~\bibnamefont
  {Akimov}}, \bibinfo {author} {\bibfnamefont {A.}~\bibnamefont {Mukherjee}},
  \bibinfo {author} {\bibfnamefont {C.~L.}\ \bibnamefont {Yu}}, \bibinfo
  {author} {\bibfnamefont {D.~E.}\ \bibnamefont {Chang}}, \bibinfo {author}
  {\bibfnamefont {A.~S.}\ \bibnamefont {Zibrov}}, \bibinfo {author}
  {\bibfnamefont {P.~R.}\ \bibnamefont {Hemmer}}, \bibinfo {author}
  {\bibfnamefont {H.}~\bibnamefont {Park}}, \ and\ \bibinfo {author}
  {\bibfnamefont {M.~D.}\ \bibnamefont {Lukin}},\ }\href@noop {} {\bibfield
  {journal} {\bibinfo  {journal} {Nature}\ }\textbf {\bibinfo {volume} {450}},\
  \bibinfo {pages} {402} (\bibinfo {year} {2007})}\BibitemShut {NoStop}%
\bibitem [{\citenamefont {Claudon}\ \emph {et~al.}(2010)\citenamefont
  {Claudon}, \citenamefont {Bleuse}, \citenamefont {Malik}, \citenamefont
  {Bazin}, \citenamefont {Jaffrennou}, \citenamefont {Gregersen}, \citenamefont
  {Sauvan}, \citenamefont {Lalanne},\ and\ \citenamefont
  {G{\'e}rard}}]{Claudon_NatPhot_10}%
  \BibitemOpen
  \bibfield  {author} {\bibinfo {author} {\bibfnamefont {J.}~\bibnamefont
  {Claudon}}, \bibinfo {author} {\bibfnamefont {J.}~\bibnamefont {Bleuse}},
  \bibinfo {author} {\bibfnamefont {N.~S.}\ \bibnamefont {Malik}}, \bibinfo
  {author} {\bibfnamefont {M.}~\bibnamefont {Bazin}}, \bibinfo {author}
  {\bibfnamefont {P.}~\bibnamefont {Jaffrennou}}, \bibinfo {author}
  {\bibfnamefont {N.}~\bibnamefont {Gregersen}}, \bibinfo {author}
  {\bibfnamefont {C.}~\bibnamefont {Sauvan}}, \bibinfo {author} {\bibfnamefont
  {P.}~\bibnamefont {Lalanne}}, \ and\ \bibinfo {author} {\bibfnamefont
  {J.-M.}\ \bibnamefont {G{\'e}rard}},\ }\href@noop {} {\bibfield  {journal}
  {\bibinfo  {journal} {Nature Photon.}\ }\textbf {\bibinfo {volume} {4}},\
  \bibinfo {pages} {174} (\bibinfo {year} {2010})}\BibitemShut {NoStop}%
\bibitem [{\citenamefont {Babinec}\ \emph {et~al.}(2010)\citenamefont
  {Babinec}, \citenamefont {Hausmann}, \citenamefont {Khan}, \citenamefont
  {Zhang}, \citenamefont {Maze}, \citenamefont {Hemmer},\ and\ \citenamefont
  {Lon\v{c}ar}}]{Babinec_NatNano_10}%
  \BibitemOpen
  \bibfield  {author} {\bibinfo {author} {\bibfnamefont {T.~M.}\ \bibnamefont
  {Babinec}}, \bibinfo {author} {\bibfnamefont {B.~M.}\ \bibnamefont
  {Hausmann}}, \bibinfo {author} {\bibfnamefont {M.}~\bibnamefont {Khan}},
  \bibinfo {author} {\bibfnamefont {Y.}~\bibnamefont {Zhang}}, \bibinfo
  {author} {\bibfnamefont {J.~R.}\ \bibnamefont {Maze}}, \bibinfo {author}
  {\bibfnamefont {P.~R.}\ \bibnamefont {Hemmer}}, \ and\ \bibinfo {author}
  {\bibfnamefont {M.}~\bibnamefont {Lon\v{c}ar}},\ }\href@noop {} {\bibfield
  {journal} {\bibinfo  {journal} {Nature Nanotech.}\ }\textbf {\bibinfo
  {volume} {5}},\ \bibinfo {pages} {195} (\bibinfo {year} {2010})}\BibitemShut
  {NoStop}%
\bibitem [{\citenamefont {G{\'e}rard}(2003)}]{Gerard_TAP_03}%
  \BibitemOpen
  \bibfield  {author} {\bibinfo {author} {\bibfnamefont {J.~M.}\ \bibnamefont
  {G{\'e}rard}},\ }\href@noop {} {\bibfield  {journal} {\bibinfo  {journal}
  {Top. Appl. Phys.}\ }\textbf {\bibinfo {volume} {90}},\ \bibinfo {pages}
  {269} (\bibinfo {year} {2003})}\BibitemShut {NoStop}%
\bibitem [{\citenamefont {Vahala}(2003)}]{Vahala_Nature_03}%
  \BibitemOpen
  \bibfield  {author} {\bibinfo {author} {\bibfnamefont {K.~J.}\ \bibnamefont
  {Vahala}},\ }\href@noop {} {\bibfield  {journal} {\bibinfo  {journal}
  {Nature}\ }\textbf {\bibinfo {volume} {423}},\ \bibinfo {pages} {839}
  (\bibinfo {year} {2003})}\BibitemShut {NoStop}%
\bibitem [{\citenamefont {Reithmaier}(2008)}]{Reithmaier_SST_08}%
  \BibitemOpen
  \bibfield  {author} {\bibinfo {author} {\bibfnamefont {J.~P.}\ \bibnamefont
  {Reithmaier}},\ }\href@noop {} {\bibfield  {journal} {\bibinfo  {journal}
  {Semicond. Sci. Technol.}\ }\textbf {\bibinfo {volume} {23}},\ \bibinfo
  {pages} {123001} (\bibinfo {year} {2008})}\BibitemShut {NoStop}%
\bibitem [{\citenamefont {Chang}\ \emph {et~al.}(2007)\citenamefont {Chang},
  \citenamefont {S{\o}rensen}, \citenamefont {Demler},\ and\ \citenamefont
  {Lukin}}]{Chang_NatPhys_07}%
  \BibitemOpen
  \bibfield  {author} {\bibinfo {author} {\bibfnamefont {D.~E.}\ \bibnamefont
  {Chang}}, \bibinfo {author} {\bibfnamefont {A.~S.}\ \bibnamefont
  {S{\o}rensen}}, \bibinfo {author} {\bibfnamefont {E.~A.}\ \bibnamefont
  {Demler}}, \ and\ \bibinfo {author} {\bibfnamefont {M.~D.}\ \bibnamefont
  {Lukin}},\ }\href@noop {} {\bibfield  {journal} {\bibinfo  {journal} {Nature
  Phys.}\ }\textbf {\bibinfo {volume} {3}},\ \bibinfo {pages} {807} (\bibinfo
  {year} {2007})}\BibitemShut {NoStop}%
\bibitem [{\citenamefont {Roy}(2011)}]{Roy_PRL_11}%
  \BibitemOpen
  \bibfield  {author} {\bibinfo {author} {\bibfnamefont {D.}~\bibnamefont
  {Roy}},\ }\href@noop {} {\bibfield  {journal} {\bibinfo  {journal} {Phys.
  Rev. Lett.}\ }\textbf {\bibinfo {volume} {106}},\ \bibinfo {pages} {053601}
  (\bibinfo {year} {2011})}\BibitemShut {NoStop}%
\bibitem [{\citenamefont {Zheng}\ \emph {et~al.}(2011)\citenamefont {Zheng},
  \citenamefont {Gauthier},\ and\ \citenamefont {Baranger}}]{Zheng_PRL_11}%
  \BibitemOpen
  \bibfield  {author} {\bibinfo {author} {\bibfnamefont {H.}~\bibnamefont
  {Zheng}}, \bibinfo {author} {\bibfnamefont {D.~J.}\ \bibnamefont {Gauthier}},
  \ and\ \bibinfo {author} {\bibfnamefont {H.~U.}\ \bibnamefont {Baranger}},\
  }\href@noop {} {\bibfield  {journal} {\bibinfo  {journal} {Phys. Rev. Lett.}\
  }\textbf {\bibinfo {volume} {107}},\ \bibinfo {pages} {223601} (\bibinfo
  {year} {2011})}\BibitemShut {NoStop}%
\bibitem [{\citenamefont {Kolchin}\ \emph {et~al.}(2011)\citenamefont
  {Kolchin}, \citenamefont {Oulton},\ and\ \citenamefont
  {Zhang}}]{Kolchin_PRL_11}%
  \BibitemOpen
  \bibfield  {author} {\bibinfo {author} {\bibfnamefont {P.}~\bibnamefont
  {Kolchin}}, \bibinfo {author} {\bibfnamefont {R.~F.}\ \bibnamefont {Oulton}},
  \ and\ \bibinfo {author} {\bibfnamefont {X.}~\bibnamefont {Zhang}},\
  }\href@noop {} {\bibfield  {journal} {\bibinfo  {journal} {Phys. Rev. Lett.}\
  }\textbf {\bibinfo {volume} {106}},\ \bibinfo {pages} {113601} (\bibinfo
  {year} {2011})}\BibitemShut {NoStop}%
\bibitem [{\citenamefont {Valente}\ \emph {et~al.}(2012)\citenamefont
  {Valente}, \citenamefont {Li}, \citenamefont {Poizat}, \citenamefont
  {G{\'e}rard}, \citenamefont {Kwek}, \citenamefont {Santos},\ and\
  \citenamefont {Auff{\`e}ves}}]{Valente_PRA_12}%
  \BibitemOpen
  \bibfield  {author} {\bibinfo {author} {\bibfnamefont {D.}~\bibnamefont
  {Valente}}, \bibinfo {author} {\bibfnamefont {Y.}~\bibnamefont {Li}},
  \bibinfo {author} {\bibfnamefont {J.~P.}\ \bibnamefont {Poizat}}, \bibinfo
  {author} {\bibfnamefont {J.~M.}\ \bibnamefont {G{\'e}rard}}, \bibinfo
  {author} {\bibfnamefont {L.~C.}\ \bibnamefont {Kwek}}, \bibinfo {author}
  {\bibfnamefont {M.~F.}\ \bibnamefont {Santos}}, \ and\ \bibinfo {author}
  {\bibfnamefont {A.}~\bibnamefont {Auff{\`e}ves}},\ }\href@noop {} {\bibfield
  {journal} {\bibinfo  {journal} {Phys. Rev. A}\ }\textbf {\bibinfo {volume}
  {86}},\ \bibinfo {pages} {022333} (\bibinfo {year} {2012})}\BibitemShut
  {NoStop}%
\bibitem [{\citenamefont {Heinrich}\ \emph {et~al.}(2010)\citenamefont
  {Heinrich}, \citenamefont {Huggenberger}, \citenamefont {Heindel},
  \citenamefont {Reitzenstein}, \citenamefont {H{\"o}fling}, \citenamefont
  {Worschech},\ and\ \citenamefont {Forchel}}]{Heinrich_APL_10}%
  \BibitemOpen
  \bibfield  {author} {\bibinfo {author} {\bibfnamefont {J.}~\bibnamefont
  {Heinrich}}, \bibinfo {author} {\bibfnamefont {A.}~\bibnamefont
  {Huggenberger}}, \bibinfo {author} {\bibfnamefont {T.}~\bibnamefont
  {Heindel}}, \bibinfo {author} {\bibfnamefont {S.}~\bibnamefont
  {Reitzenstein}}, \bibinfo {author} {\bibfnamefont {S.}~\bibnamefont
  {H{\"o}fling}}, \bibinfo {author} {\bibfnamefont {L.}~\bibnamefont
  {Worschech}}, \ and\ \bibinfo {author} {\bibfnamefont {A.}~\bibnamefont
  {Forchel}},\ }\href@noop {} {\bibfield  {journal} {\bibinfo  {journal} {Appl.
  Phys. Lett.}\ }\textbf {\bibinfo {volume} {96}},\ \bibinfo {pages} {211117}
  (\bibinfo {year} {2010})}\BibitemShut {NoStop}%
\bibitem [{\citenamefont {Reimer}\ \emph {et~al.}(2012)\citenamefont {Reimer},
  \citenamefont {Bulgarini}, \citenamefont {Akopian}, \citenamefont {Hocevar},
  \citenamefont {Bavinck}, \citenamefont {Verheijen}, \citenamefont {Bakkers},
  \citenamefont {Kouwenhoven},\ and\ \citenamefont
  {Zwiller}}]{Reimer_NatComm_11}%
  \BibitemOpen
  \bibfield  {author} {\bibinfo {author} {\bibfnamefont {M.~E.}\ \bibnamefont
  {Reimer}}, \bibinfo {author} {\bibfnamefont {G.}~\bibnamefont {Bulgarini}},
  \bibinfo {author} {\bibfnamefont {N.}~\bibnamefont {Akopian}}, \bibinfo
  {author} {\bibfnamefont {M.}~\bibnamefont {Hocevar}}, \bibinfo {author}
  {\bibfnamefont {M.~B.}\ \bibnamefont {Bavinck}}, \bibinfo {author}
  {\bibfnamefont {M.~A.}\ \bibnamefont {Verheijen}}, \bibinfo {author}
  {\bibfnamefont {E.~P.}\ \bibnamefont {Bakkers}}, \bibinfo {author}
  {\bibfnamefont {L.~P.}\ \bibnamefont {Kouwenhoven}}, \ and\ \bibinfo {author}
  {\bibfnamefont {V.}~\bibnamefont {Zwiller}},\ }\href@noop {} {\bibfield
  {journal} {\bibinfo  {journal} {Nature Comm.}\ }\textbf {\bibinfo {volume}
  {3}},\ \bibinfo {pages} {737} (\bibinfo {year} {2012})}\BibitemShut {NoStop}%
\bibitem [{\citenamefont {Bleuse}\ \emph {et~al.}(2011)\citenamefont {Bleuse},
  \citenamefont {Claudon}, \citenamefont {Creasey}, \citenamefont {Malik},
  \citenamefont {G\'erard}, \citenamefont {Maksymov}, \citenamefont {Hugonin},\
  and\ \citenamefont {Lalanne}}]{Bleuse_PRL_11}%
  \BibitemOpen
  \bibfield  {author} {\bibinfo {author} {\bibfnamefont {J.}~\bibnamefont
  {Bleuse}}, \bibinfo {author} {\bibfnamefont {J.}~\bibnamefont {Claudon}},
  \bibinfo {author} {\bibfnamefont {M.}~\bibnamefont {Creasey}}, \bibinfo
  {author} {\bibfnamefont {N.~S.}\ \bibnamefont {Malik}}, \bibinfo {author}
  {\bibfnamefont {J.-M.}\ \bibnamefont {G\'erard}}, \bibinfo {author}
  {\bibfnamefont {I.}~\bibnamefont {Maksymov}}, \bibinfo {author}
  {\bibfnamefont {J.-P.}\ \bibnamefont {Hugonin}}, \ and\ \bibinfo {author}
  {\bibfnamefont {P.}~\bibnamefont {Lalanne}},\ }\href@noop {} {\bibfield
  {journal} {\bibinfo  {journal} {Phys. Rev. Lett.}\ }\textbf {\bibinfo
  {volume} {106}},\ \bibinfo {pages} {103601} (\bibinfo {year}
  {2011})}\BibitemShut {NoStop}%
\bibitem [{\citenamefont {Munsch}\ \emph {et~al.}(2012)\citenamefont {Munsch},
  \citenamefont {Claudon}, \citenamefont {Bleuse}, \citenamefont {Malik},
  \citenamefont {Dupuy}, \citenamefont {G{\'e}rard}, \citenamefont {Chen},
  \citenamefont {Gregersen},\ and\ \citenamefont {M{\o}rk}}]{Munsch_PRL_12}%
  \BibitemOpen
  \bibfield  {author} {\bibinfo {author} {\bibfnamefont {M.}~\bibnamefont
  {Munsch}}, \bibinfo {author} {\bibfnamefont {J.}~\bibnamefont {Claudon}},
  \bibinfo {author} {\bibfnamefont {J.}~\bibnamefont {Bleuse}}, \bibinfo
  {author} {\bibfnamefont {N.~S.}\ \bibnamefont {Malik}}, \bibinfo {author}
  {\bibfnamefont {E.}~\bibnamefont {Dupuy}}, \bibinfo {author} {\bibfnamefont
  {J.-M.}\ \bibnamefont {G{\'e}rard}}, \bibinfo {author} {\bibfnamefont
  {Y.}~\bibnamefont {Chen}}, \bibinfo {author} {\bibfnamefont {N.}~\bibnamefont
  {Gregersen}}, \ and\ \bibinfo {author} {\bibfnamefont {J.}~\bibnamefont
  {M{\o}rk}},\ }\href@noop {} {\bibfield  {journal} {\bibinfo  {journal} {Phys.
  Rev. Lett.}\ }\textbf {\bibinfo {volume} {108}},\ \bibinfo {pages} {077405}
  (\bibinfo {year} {2012})}\BibitemShut {NoStop}%
\bibitem [{\citenamefont {Bulgarini}\ \emph {et~al.}(2012)\citenamefont
  {Bulgarini}, \citenamefont {Reimer}, \citenamefont {Zehender}, \citenamefont
  {Hocevar}, \citenamefont {Bakkers}, \citenamefont {Kouwenhoven},\ and\
  \citenamefont {Zwiller}}]{Bulgarini_APL_12}%
  \BibitemOpen
  \bibfield  {author} {\bibinfo {author} {\bibfnamefont {G.}~\bibnamefont
  {Bulgarini}}, \bibinfo {author} {\bibfnamefont {M.~E.}\ \bibnamefont
  {Reimer}}, \bibinfo {author} {\bibfnamefont {T.}~\bibnamefont {Zehender}},
  \bibinfo {author} {\bibfnamefont {M.}~\bibnamefont {Hocevar}}, \bibinfo
  {author} {\bibfnamefont {E.~P. A.~M.}\ \bibnamefont {Bakkers}}, \bibinfo
  {author} {\bibfnamefont {L.~P.}\ \bibnamefont {Kouwenhoven}}, \ and\ \bibinfo
  {author} {\bibfnamefont {V.}~\bibnamefont {Zwiller}},\ }\href@noop {}
  {\bibfield  {journal} {\bibinfo  {journal} {Appl. Phys. Lett.}\ }\textbf
  {\bibinfo {volume} {100}},\ \bibinfo {pages} {121106} (\bibinfo {year}
  {2012})}\BibitemShut {NoStop}%
\bibitem [{\citenamefont {Friedler}\ \emph {et~al.}(2008)\citenamefont
  {Friedler}, \citenamefont {Lalanne}, \citenamefont {Hugonin}, \citenamefont
  {Claudon}, \citenamefont {G{\'e}rard}, \citenamefont {Beveratos},\ and\
  \citenamefont {Robert-Philip}}]{Friedler_OL_08}%
  \BibitemOpen
  \bibfield  {author} {\bibinfo {author} {\bibfnamefont {I.}~\bibnamefont
  {Friedler}}, \bibinfo {author} {\bibfnamefont {P.}~\bibnamefont {Lalanne}},
  \bibinfo {author} {\bibfnamefont {J.~P.}\ \bibnamefont {Hugonin}}, \bibinfo
  {author} {\bibfnamefont {J.}~\bibnamefont {Claudon}}, \bibinfo {author}
  {\bibfnamefont {J.~M.}\ \bibnamefont {G{\'e}rard}}, \bibinfo {author}
  {\bibfnamefont {A.}~\bibnamefont {Beveratos}}, \ and\ \bibinfo {author}
  {\bibfnamefont {I.}~\bibnamefont {Robert-Philip}},\ }\href@noop {} {\bibfield
   {journal} {\bibinfo  {journal} {Opt. Lett.}\ }\textbf {\bibinfo {volume}
  {33}},\ \bibinfo {pages} {2635} (\bibinfo {year} {2008})}\BibitemShut
  {NoStop}%
\bibitem [{\citenamefont {Gregersen}\ \emph {et~al.}(2008)\citenamefont
  {Gregersen}, \citenamefont {Nielsen}, \citenamefont {Claudon}, \citenamefont
  {G{\'e}rard},\ and\ \citenamefont {M{\o}rk}}]{Gregersen_OL_08}%
  \BibitemOpen
  \bibfield  {author} {\bibinfo {author} {\bibfnamefont {N.}~\bibnamefont
  {Gregersen}}, \bibinfo {author} {\bibfnamefont {T.~R.}\ \bibnamefont
  {Nielsen}}, \bibinfo {author} {\bibfnamefont {J.}~\bibnamefont {Claudon}},
  \bibinfo {author} {\bibfnamefont {J.~M.}\ \bibnamefont {G{\'e}rard}}, \ and\
  \bibinfo {author} {\bibfnamefont {J.}~\bibnamefont {M{\o}rk}},\ }\href@noop
  {} {\bibfield  {journal} {\bibinfo  {journal} {Opt. Lett.}\ }\textbf
  {\bibinfo {volume} {33}},\ \bibinfo {pages} {1693} (\bibinfo {year}
  {2008})}\BibitemShut {NoStop}%
\bibitem [{\citenamefont {Friedler}\ \emph {et~al.}(2009)\citenamefont
  {Friedler}, \citenamefont {Sauvan}, \citenamefont {Hugonin}, \citenamefont
  {Lalanne}, \citenamefont {Claudon},\ and\ \citenamefont
  {G{\'e}rard}}]{Friedler_OE_09}%
  \BibitemOpen
  \bibfield  {author} {\bibinfo {author} {\bibfnamefont {I.}~\bibnamefont
  {Friedler}}, \bibinfo {author} {\bibfnamefont {C.}~\bibnamefont {Sauvan}},
  \bibinfo {author} {\bibfnamefont {J.~P.}\ \bibnamefont {Hugonin}}, \bibinfo
  {author} {\bibfnamefont {P.}~\bibnamefont {Lalanne}}, \bibinfo {author}
  {\bibfnamefont {J.}~\bibnamefont {Claudon}}, \ and\ \bibinfo {author}
  {\bibfnamefont {J.~M.}\ \bibnamefont {G{\'e}rard}},\ }\href@noop {}
  {\bibfield  {journal} {\bibinfo  {journal} {Opt. Express}\ }\textbf {\bibinfo
  {volume} {17}},\ \bibinfo {pages} {2095} (\bibinfo {year}
  {2009})}\BibitemShut {NoStop}%
\bibitem [{\citenamefont {Rakher}\ \emph {et~al.}(2009)\citenamefont {Rakher},
  \citenamefont {Stoltz}, \citenamefont {Coldren}, \citenamefont {Petroff},\
  and\ \citenamefont {Bouwmeester}}]{Rakher_PRL_09}%
  \BibitemOpen
  \bibfield  {author} {\bibinfo {author} {\bibfnamefont {M.~T.}\ \bibnamefont
  {Rakher}}, \bibinfo {author} {\bibfnamefont {N.}~\bibnamefont {Stoltz}},
  \bibinfo {author} {\bibfnamefont {L.~A.}\ \bibnamefont {Coldren}}, \bibinfo
  {author} {\bibfnamefont {P.~M.}\ \bibnamefont {Petroff}}, \ and\ \bibinfo
  {author} {\bibfnamefont {D.}~\bibnamefont {Bouwmeester}},\ }\href@noop {}
  {\bibfield  {journal} {\bibinfo  {journal} {Phys. Rev. Lett.}\ }\textbf
  {\bibinfo {volume} {102}},\ \bibinfo {pages} {097403} (\bibinfo {year}
  {2009})}\BibitemShut {NoStop}%
\bibitem [{\citenamefont {Kimble}(2008)}]{Kimble_Nature_08}%
  \BibitemOpen
  \bibfield  {author} {\bibinfo {author} {\bibfnamefont {H.~J.}\ \bibnamefont
  {Kimble}},\ }\href@noop {} {\bibfield  {journal} {\bibinfo  {journal}
  {Nature}\ }\textbf {\bibinfo {volume} {453}},\ \bibinfo {pages} {1023}
  (\bibinfo {year} {2008})}\BibitemShut {NoStop}%
\bibitem [{\citenamefont {Bienstman}\ and\ \citenamefont
  {Baets}(2001)}]{Bienstman_OQE_01}%
  \BibitemOpen
  \bibfield  {author} {\bibinfo {author} {\bibfnamefont {P.}~\bibnamefont
  {Bienstman}}\ and\ \bibinfo {author} {\bibfnamefont {R.}~\bibnamefont
  {Baets}},\ }\href@noop {} {\bibfield  {journal} {\bibinfo  {journal} {Opt.
  Quantum Electron.}\ }\textbf {\bibinfo {volume} {33}},\ \bibinfo {pages}
  {327} (\bibinfo {year} {2001})}\BibitemShut {NoStop}%
\bibitem [{\citenamefont {Gregersen}\ and\ \citenamefont
  {M{\o}rk}(2008)}]{Gregersen_OQE_08}%
  \BibitemOpen
  \bibfield  {author} {\bibinfo {author} {\bibfnamefont {N.}~\bibnamefont
  {Gregersen}}\ and\ \bibinfo {author} {\bibfnamefont {J.}~\bibnamefont
  {M{\o}rk}},\ }\href@noop {} {\bibfield  {journal} {\bibinfo  {journal} {Opt.
  Quantum Electron.}\ }\textbf {\bibinfo {volume} {40}},\ \bibinfo {pages}
  {957} (\bibinfo {year} {2008})}\BibitemShut {NoStop}%
\bibitem [{\citenamefont {Pelton}\ \emph {et~al.}(2002)\citenamefont {Pelton},
  \citenamefont {Santori}, \citenamefont {Vuckovic}, \citenamefont {Zhang},
  \citenamefont {Solomon}, \citenamefont {Plant},\ and\ \citenamefont
  {Yamamoto}}]{Pelton_PRL_02}%
  \BibitemOpen
  \bibfield  {author} {\bibinfo {author} {\bibfnamefont {M.}~\bibnamefont
  {Pelton}}, \bibinfo {author} {\bibfnamefont {C.}~\bibnamefont {Santori}},
  \bibinfo {author} {\bibfnamefont {J.}~\bibnamefont {Vuckovic}}, \bibinfo
  {author} {\bibfnamefont {B.}~\bibnamefont {Zhang}}, \bibinfo {author}
  {\bibfnamefont {G.~S.}\ \bibnamefont {Solomon}}, \bibinfo {author}
  {\bibfnamefont {J.}~\bibnamefont {Plant}}, \ and\ \bibinfo {author}
  {\bibfnamefont {Y.}~\bibnamefont {Yamamoto}},\ }\href@noop {} {\bibfield
  {journal} {\bibinfo  {journal} {Phys. Rev. Lett.}\ }\textbf {\bibinfo
  {volume} {89}},\ \bibinfo {pages} {233602} (\bibinfo {year}
  {2002})}\BibitemShut {NoStop}%
\bibitem [{\citenamefont {Strauf}\ \emph {et~al.}(2007)\citenamefont {Strauf},
  \citenamefont {Stoltz}, \citenamefont {Rakher}, \citenamefont {Coldren},
  \citenamefont {Petroff},\ and\ \citenamefont
  {Bouwmeester}}]{Strauf_NatPhot_07}%
  \BibitemOpen
  \bibfield  {author} {\bibinfo {author} {\bibfnamefont {S.}~\bibnamefont
  {Strauf}}, \bibinfo {author} {\bibfnamefont {N.~G.}\ \bibnamefont {Stoltz}},
  \bibinfo {author} {\bibfnamefont {M.~T.}\ \bibnamefont {Rakher}}, \bibinfo
  {author} {\bibfnamefont {L.~A.}\ \bibnamefont {Coldren}}, \bibinfo {author}
  {\bibfnamefont {P.~M.}\ \bibnamefont {Petroff}}, \ and\ \bibinfo {author}
  {\bibfnamefont {D.}~\bibnamefont {Bouwmeester}},\ }\href@noop {} {\bibfield
  {journal} {\bibinfo  {journal} {Nature Photon.}\ }\textbf {\bibinfo {volume}
  {1}},\ \bibinfo {pages} {704} (\bibinfo {year} {2007})}\BibitemShut {NoStop}%
\bibitem [{\citenamefont {Lee}\ \emph {et~al.}(2011)\citenamefont {Lee},
  \citenamefont {Chen}, \citenamefont {Eghlidi}, \citenamefont {Kukura},
  \citenamefont {Lettow}, \citenamefont {Renn}, \citenamefont {Sandoghdar},\
  and\ \citenamefont {G{\"o}tzinger}}]{Lee_NatPhot_11}%
  \BibitemOpen
  \bibfield  {author} {\bibinfo {author} {\bibfnamefont {K.~G.}\ \bibnamefont
  {Lee}}, \bibinfo {author} {\bibfnamefont {X.~W.}\ \bibnamefont {Chen}},
  \bibinfo {author} {\bibfnamefont {H.}~\bibnamefont {Eghlidi}}, \bibinfo
  {author} {\bibfnamefont {P.}~\bibnamefont {Kukura}}, \bibinfo {author}
  {\bibfnamefont {R.}~\bibnamefont {Lettow}}, \bibinfo {author} {\bibfnamefont
  {A.}~\bibnamefont {Renn}}, \bibinfo {author} {\bibfnamefont {V.}~\bibnamefont
  {Sandoghdar}}, \ and\ \bibinfo {author} {\bibfnamefont {S.}~\bibnamefont
  {G{\"o}tzinger}},\ }\href@noop {} {\bibfield  {journal} {\bibinfo  {journal}
  {Nature Photon.}\ }\textbf {\bibinfo {volume} {5}},\ \bibinfo {pages} {166}
  (\bibinfo {year} {2011})}\BibitemShut {NoStop}%
\bibitem [{\citenamefont {B{\"o}ckler}\ \emph {et~al.}(2008)\citenamefont
  {B{\"o}ckler}, \citenamefont {Reitzenstein}, \citenamefont {Kistner},
  \citenamefont {Debusmann}, \citenamefont {L{\"o}ffler}, \citenamefont {Kida},
  \citenamefont {H{\"o}fling}, \citenamefont {Forchel}, \citenamefont
  {Grenouillet}, \citenamefont {Claudon},\ and\ \citenamefont
  {G{\'e}rard}}]{Bockler_APL_08}%
  \BibitemOpen
  \bibfield  {author} {\bibinfo {author} {\bibfnamefont {C.}~\bibnamefont
  {B{\"o}ckler}}, \bibinfo {author} {\bibfnamefont {S.}~\bibnamefont
  {Reitzenstein}}, \bibinfo {author} {\bibfnamefont {C.}~\bibnamefont
  {Kistner}}, \bibinfo {author} {\bibfnamefont {R.}~\bibnamefont {Debusmann}},
  \bibinfo {author} {\bibfnamefont {A.}~\bibnamefont {L{\"o}ffler}}, \bibinfo
  {author} {\bibfnamefont {T.}~\bibnamefont {Kida}}, \bibinfo {author}
  {\bibfnamefont {S.}~\bibnamefont {H{\"o}fling}}, \bibinfo {author}
  {\bibfnamefont {A.}~\bibnamefont {Forchel}}, \bibinfo {author} {\bibfnamefont
  {L.}~\bibnamefont {Grenouillet}}, \bibinfo {author} {\bibfnamefont
  {J.}~\bibnamefont {Claudon}}, \ and\ \bibinfo {author} {\bibfnamefont
  {J.~M.}\ \bibnamefont {G{\'e}rard}},\ }\href@noop {} {\bibfield  {journal}
  {\bibinfo  {journal} {Appl. Phys. Lett.}\ }\textbf {\bibinfo {volume} {92}},\
  \bibinfo {pages} {091107} (\bibinfo {year} {2008})}\BibitemShut {NoStop}%
\bibitem [{\citenamefont {Gregersen}\ \emph {et~al.}(2010)\citenamefont
  {Gregersen}, \citenamefont {Nielsen}, \citenamefont {M{\o}rk}, \citenamefont
  {Claudon},\ and\ \citenamefont {G{\'e}rard}}]{Gregersen_OE_10}%
  \BibitemOpen
  \bibfield  {author} {\bibinfo {author} {\bibfnamefont {N.}~\bibnamefont
  {Gregersen}}, \bibinfo {author} {\bibfnamefont {T.~R.}\ \bibnamefont
  {Nielsen}}, \bibinfo {author} {\bibfnamefont {J.}~\bibnamefont {M{\o}rk}},
  \bibinfo {author} {\bibfnamefont {J.}~\bibnamefont {Claudon}}, \ and\
  \bibinfo {author} {\bibfnamefont {J.-M.}\ \bibnamefont {G{\'e}rard}},\
  }\href@noop {} {\bibfield  {journal} {\bibinfo  {journal} {Opt. Express}\
  }\textbf {\bibinfo {volume} {18}},\ \bibinfo {pages} {21204} (\bibinfo {year}
  {2010})}\BibitemShut {NoStop}%
\bibitem [{\citenamefont {Yuan}\ \emph {et~al.}(2002)\citenamefont {Yuan},
  \citenamefont {Kardynal}, \citenamefont {Stevenson}, \citenamefont {Shields},
  \citenamefont {Lobo}, \citenamefont {Cooper}, \citenamefont {Beattie},
  \citenamefont {Ritchie},\ and\ \citenamefont {Pepper}}]{Yuan_Science_02}%
  \BibitemOpen
  \bibfield  {author} {\bibinfo {author} {\bibfnamefont {Z.}~\bibnamefont
  {Yuan}}, \bibinfo {author} {\bibfnamefont {B.}~\bibnamefont {Kardynal}},
  \bibinfo {author} {\bibfnamefont {R.}~\bibnamefont {Stevenson}}, \bibinfo
  {author} {\bibfnamefont {A.}~\bibnamefont {Shields}}, \bibinfo {author}
  {\bibfnamefont {C.}~\bibnamefont {Lobo}}, \bibinfo {author} {\bibfnamefont
  {K.}~\bibnamefont {Cooper}}, \bibinfo {author} {\bibfnamefont
  {N.}~\bibnamefont {Beattie}}, \bibinfo {author} {\bibfnamefont
  {D.}~\bibnamefont {Ritchie}}, \ and\ \bibinfo {author} {\bibfnamefont
  {M.}~\bibnamefont {Pepper}},\ }\href@noop {} {\bibfield  {journal} {\bibinfo
  {journal} {Science}\ }\textbf {\bibinfo {volume} {295}},\ \bibinfo {pages}
  {102} (\bibinfo {year} {2002})}\BibitemShut {NoStop}%
\bibitem [{\citenamefont {Finley}\ \emph {et~al.}(2004)\citenamefont {Finley},
  \citenamefont {Sabathil}, \citenamefont {Vogl}, \citenamefont {Abstreiter},
  \citenamefont {Oulton}, \citenamefont {Tartakovskii}, \citenamefont
  {Mowbray}, \citenamefont {Skolnick}, \citenamefont {Liew}, \citenamefont
  {Cullis},\ and\ \citenamefont {Hopkinson}}]{Finley_PRB(R)_04}%
  \BibitemOpen
  \bibfield  {author} {\bibinfo {author} {\bibfnamefont {J.~J.}\ \bibnamefont
  {Finley}}, \bibinfo {author} {\bibfnamefont {M.}~\bibnamefont {Sabathil}},
  \bibinfo {author} {\bibfnamefont {P.}~\bibnamefont {Vogl}}, \bibinfo {author}
  {\bibfnamefont {G.}~\bibnamefont {Abstreiter}}, \bibinfo {author}
  {\bibfnamefont {R.}~\bibnamefont {Oulton}}, \bibinfo {author} {\bibfnamefont
  {A.~I.}\ \bibnamefont {Tartakovskii}}, \bibinfo {author} {\bibfnamefont
  {D.~J.}\ \bibnamefont {Mowbray}}, \bibinfo {author} {\bibfnamefont {M.~S.}\
  \bibnamefont {Skolnick}}, \bibinfo {author} {\bibfnamefont {S.~L.}\
  \bibnamefont {Liew}}, \bibinfo {author} {\bibfnamefont {A.~G.}\ \bibnamefont
  {Cullis}}, \ and\ \bibinfo {author} {\bibfnamefont {M.}~\bibnamefont
  {Hopkinson}},\ }\href@noop {} {\bibfield  {journal} {\bibinfo  {journal}
  {Phys. Rev. B}\ }\textbf {\bibinfo {volume} {70}},\ \bibinfo {pages}
  {201308(R)} (\bibinfo {year} {2004})}\BibitemShut {NoStop}%
\bibitem [{\citenamefont {Bennett}\ \emph {et~al.}(2010)\citenamefont
  {Bennett}, \citenamefont {Pooley}, \citenamefont {Stevenson}, \citenamefont
  {Ward}, \citenamefont {Patel}, \citenamefont {de~la Giroday}, \citenamefont
  {Sk{\"o}ld}, \citenamefont {Farrer}, \citenamefont {Nicoll}, \citenamefont
  {Ritchie},\ and\ \citenamefont {Shields}}]{Bennett_NatPhys_10}%
  \BibitemOpen
  \bibfield  {author} {\bibinfo {author} {\bibfnamefont {A.~J.}\ \bibnamefont
  {Bennett}}, \bibinfo {author} {\bibfnamefont {M.~A.}\ \bibnamefont {Pooley}},
  \bibinfo {author} {\bibfnamefont {R.~M.}\ \bibnamefont {Stevenson}}, \bibinfo
  {author} {\bibfnamefont {M.~B.}\ \bibnamefont {Ward}}, \bibinfo {author}
  {\bibfnamefont {R.~B.}\ \bibnamefont {Patel}}, \bibinfo {author}
  {\bibfnamefont {A.~B.}\ \bibnamefont {de~la Giroday}}, \bibinfo {author}
  {\bibfnamefont {N.}~\bibnamefont {Sk{\"o}ld}}, \bibinfo {author}
  {\bibfnamefont {I.}~\bibnamefont {Farrer}}, \bibinfo {author} {\bibfnamefont
  {C.~A.}\ \bibnamefont {Nicoll}}, \bibinfo {author} {\bibfnamefont {D.~A.}\
  \bibnamefont {Ritchie}}, \ and\ \bibinfo {author} {\bibfnamefont {A.~J.}\
  \bibnamefont {Shields}},\ }\href@noop {} {\bibfield  {journal} {\bibinfo
  {journal} {Nature Phys.}\ }\textbf {\bibinfo {volume} {6}},\ \bibinfo {pages}
  {947} (\bibinfo {year} {2010})}\BibitemShut {NoStop}%
\end{thebibliography}

%

\end{document}